\documentclass[pdflatex,sn-mathphys-num]{sn-jnl}

\usepackage{graphicx}%
\usepackage{multirow}%
\usepackage{amsmath,amssymb,amsfonts}%
\usepackage{amsthm}%
\usepackage{mathrsfs}%
\usepackage[title]{appendix}%
\usepackage{xcolor}%
\usepackage{textcomp}%
\usepackage{manyfoot}%
\usepackage{booktabs}%
\usepackage{algorithm}%
\usepackage{algorithmicx}%
\usepackage{algpseudocode}%
\usepackage{listings}%
\usepackage{subcaption}
\usepackage{url}
\theoremstyle{thmstyleone}

\theoremstyle{thmstyletwo}

\theoremstyle{thmstylethree}

\begin{document}

\title[Chaos-Enhanced Prototypical Networks]{Chaos-Enhanced Prototypical Networks for Few-Shot Medical Image Classification}

%% Authors
\author*[1]{\fnm{Chinthakuntla} \sur{Meghan Sai}}\email{meghansai6@gmail.com}

\author[1]{\fnm{Murarisetty} \sur{V Sai Kartheek}}\email{usernamekartheek@gmail.com}

\author*[1]{\fnm{Sita Devi} \sur{Bharatula}}\email{b\_sitadevi@ch.amrita.edu}

\author[2]{\fnm{Karthik} \sur{Seemakurthy}}\email{kseemakurthy@hydroniumenergies.com}

%% Affiliations
\affil*[1]{\orgdiv{Department of ECE}, \orgname{Amrita Vishwa Vidyapeetham}, 
\orgaddress{\street{Vengal}, \city{Chennai}, \postcode{601103}, \state{Tamil Nadu}, \country{India}}}

\affil[2]{\orgname{Hydronium Energies Limited}, 
\orgaddress{\street{Bristol}, \city{London}, \country{England}}}

%% Abstract
\abstract{The scarcity of labeled clinical data in oncology makes Few-Shot Learning (FSL) a critical framework for Computer Aided Diagnostics, but we observed that standard Prototypical Networks often struggle with the "prototype instability" caused by morphological noise and high intra-class variance in brain tumor scans. Our work attempts to minimize this by integrating a non-linear Logistic Chaos Module into a fine-tuned ResNet-18 backbone creating the Chaos-Enhanced ProtoNet(CE-ProtoNet). Using the deterministic ergodicity of the logistic chaos map we inject controlled perturbations into support features during episodic training-essentially for "stress testing" the embedding space. This process makes the model to converge on noise-invariant representations without increasing computational overhead. Testing this on a 4-way 5-shot brain tumor classification task, we found that a 15\% chaotic injection level worked efficiently to stabilize high-dimensional clusters and reduce class dispersion. Our method achieved a peak test accuracy of 84.52\%, outperforming standard ProtoNet. Our results suggest the idea of using chaotic perturbation as an efficient, low-overhead regularization tool, for the data-scarce regimes.}

%% Keywords

\maketitle

\section{Introduction}\label{sec1}
Deep learning has completely transformed medical imaging analysis by being adopted into clinical diagnostic pipelines. However, the performance of all these architectures is still hindered by the reliance on large (excellent) annotated datasets, which currently represent a major bottleneck in neuroradiology. Prohibitive annotation costs and stringent privacy guidelines in typical clinical environments frequently impede labeling scans for infrequent pathologies. Relying on these limited data, few-shot learning (FSL) has been an important avenue of research that enables the models to generalize from only a small number of support samples. As a solution, metric-based frameworks (especially prototypical networks (ProtoNet)) are popular for their ability to project images into a latent embedded space, where classification is performed using distance-based similarity measures.

Classical ProtoNets can perform well in generic computer vision problems. However, they are highly unstable for MRI due to their complex texture gradients. Medical pathologies often appear as fine-textured or expansionary patches in tissue density or signal intensity, whereas natural images typically exhibit sharp edges. The low contrast-to-noise ratios often lead to overlap between peritumoral edema and healthy white matter. Traditionally, a ProtoNet's class prototypes are the arithmetic mean of support features. This strategy may be easily distorted: a single supportive scan containing acquisition artifacts or an excess of intra-class variance can considerably shift the position of the prototype. Consequently, the overlapping feature space of the varied tumor morphologies limits diagnostic specificity.\cite{1}.

To deal with this situation, recent works have developed dual-channel architectures such as the CNN-Transformer to capture pathological details at multiple scales\cite{2}. Nevertheless, these solutions often double the computational cost, limiting their practical applications in clinical environments with limited resources. In the same way, during the training episodes, frameworks which require optimization and quadratic programming problems increase latency which is often unacceptable for real-time applications \ cite {3}. While stochastic regularization techniques such as Gaussian Disturbance Soft Labels \cite{4}, provides some robustness, but they often suffer with instability during convergence in multi-learner systems.

This work provides an efficient, lightweight technique using the concepts of Chaos Theory to shape the structure of embeddings. As tiny changes can lead to big differences - like the Butterfly Effect - we build unique data expansions. By few-shot learning, using a single sample image generates multiple evolving feature forms along a tight path and instead of depending on fixed snapshots, this model adapts to identify smooth shifts across wild but predictable changes. This pressure gets the embedding space closer together \cite{5}. This system works by making the model to focus on steady biological signals instead of random fluctuations. Chaotic maps behave in an unpredictable yet structured manner, they push the learning process across diverse edge cases - unlike arbitrary noise, which often lacks direction. This movement helps avoid getting stuck too early during training while also shrinking differences within the same class \cite{6,7}.	

\begin{figure}[h!]
\centering
\includegraphics[width=1.0\linewidth]{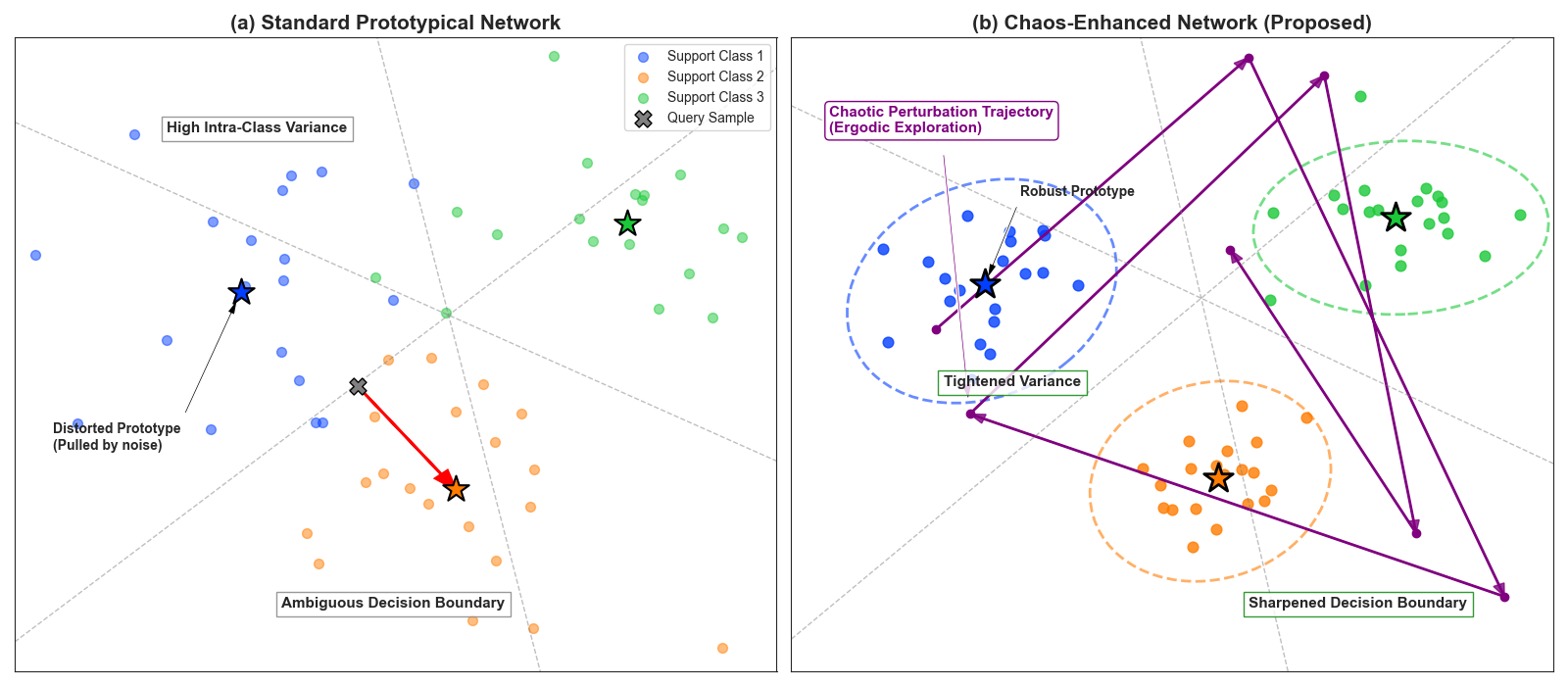}
\caption{Conceptual framework of embedding space stabilization. \textbf{(a) Standard ProtoNet:} Inherent feature noise in data-scarce regimes drives high intra-class variance, leading to overlapping clusters and displacement in centroid. \textbf{(b) Proposed CE-ProtoNet:} The integration of a Logistic Chaos Module (purple trajectory) during training works as a deterministic regularizer and forces the architecture to resolve a continuous manifold of simulated variations, thereby sharpening decision boundaries and establishing robust, noise-resilient prototypes.}
\label{fig:concept_visualization}
\end{figure}

The Chaos-Enhanced Prototypical Network (CE-ProtoNet) is trained by integrating a nonlinear Logistic Chaos Module embedded into a fine-tuned ResNet-18 backbone. Unlike the traditional static feature extractors, the Logistic Chaos Module applies deterministic jitter to shape the feature embeddings during the training process. The tumor shape variations can arise from multiple scenario's like differences in MRI Machine types or patient positions. The chaos module makes nonlinear perturbations which forces the model to adapt for superficial aspects and focus on the class's invariant features. The results of this experiment show that the 4-way, 5-shot brain tumor classification task showed that a 15\% rate of chaotic disturbances optimized the embedding space, achieving a peak test accuracy of 84.52\%. This framework provides an high-precision, parameter-efficient diagnosis in data-scarce clinical settings.

\section{Methodology}
\label{sec:methodology}

The Chaos-Enhanced Prototypical Network (CE-ProtoNet), is specifically designed to optimize the instability of few-shot medical image classification. By integrating ResNet-18 deep residual feature extraction with deterministic logistic chaotic regularization, this framework directly targets ``prototype bias'' \cite{12}. This bias arises when a restricted support set fails to capture the underlying class distribution, which severely degrades cross-patient generalization.

\subsection{Dataset and Preprocessing}
The model development was conducted using a Brain Tumor MRI dataset \cite{23} available online containing 3,264 scans distributed across four different classes: Glioma, Meningioma, Pituitary Tumor, and No Tumor. Instead of using classical static training and validation split techniques, we used pipeline that manages data through dynamic episodic sampling. In every iteration, a $N$-way $K$-shot support and query sets is created by system selecting random samples. The higher inter-class similarities between Glioma and Meningioma tissues, and significant intra-class variance due to tumor morphology make this dataset an appropriately suitable testbed for FSL.

\subsubsection{Preprocessing Pipeline}
The data is processed through a uniform preprocessing sequence to align with the ResNet-18 input requirements:
\begin{itemize}
    \item \textbf{Resizing:} Raw MRI slices are scaled to $256 \times 256$ pixels.
    \item \textbf{Center Cropping:} A $224 \times 224$ central crop isolates the brain parenchyma and discards peripheral background noise.
    \item \textbf{Normalization:} Pixel intensities are normalized using the ImageNet mean ($\mu=[0.485, 0.456, 0.406]$) and standard deviation ($\sigma=[0.229, 0.224, 0.225]$), ensuring the pre-trained convolutional filters activate correctly.
\end{itemize}

\subsection{Episodic Meta-Learning Formulation}
The training process consists of an $N$-way $K$-shot episodic protocol, making the process directly replicate test-time evaluation. An \textbf{episode} can be thought of as a single, self-contained learning task presented to the model, whereas an \textbf{iteration} refers to the step during training in which the model’s parameters are updated. In working, multiple episodes are often processed before a single iteration is completed, as the updates are typically based on the aggregated information from those episodes.

Let $\mathcal{D}_{train}$ denote the training set containing base classes $\mathcal{C}_{base}$. To build an episode, we randomly sample a subset of classes $\mathcal{C}_{episode} \subset \mathcal{C}_{base}$, where $|\mathcal{C}_{episode}| = N$. For each class $c \in \mathcal{C}_{episode}$, the algorithm creates two disjoint image sets:
\begin{itemize}
    \item \textbf{Support Set ($S$):} Contains $K$ labeled examples per class used to construct the prototype centroid. $S = \{(x_s^i, y_s^i)\}_{i=1}^{N \times K}$.
    \item \textbf{Query Set ($Q$):} Contains $Q_{count}$ samples per class used to evaluate generalization and compute the loss gradient. $Q = \{(x_q^j, y_q^j)\}_{j=1}^{N \times Q_{count}}$.
\end{itemize}

In the every iteration, the network tries to reduce the prediction error on Query Set ($Q$) based on the prototypes formulated from the Support Set ($S$). We need this episodic structure as standard static training in medical domains often produce diffused feature embeddings, creating multi class overlaps and blurring of decision boundaries \cite{10}.

\subsection{Backbone Architecture and Selective Fine-Tuning}
We used an pretrained Imagenet \textbf{ResNet-18} backbone for feature extraction as its architecture contains enough layer depth to capture complex textures while its skip connections prevent the vanishing gradients that plague deeper networks in data-scarce regimes.

\begin{figure}[h!]
    \centering
    \includegraphics[width=1.0\linewidth]{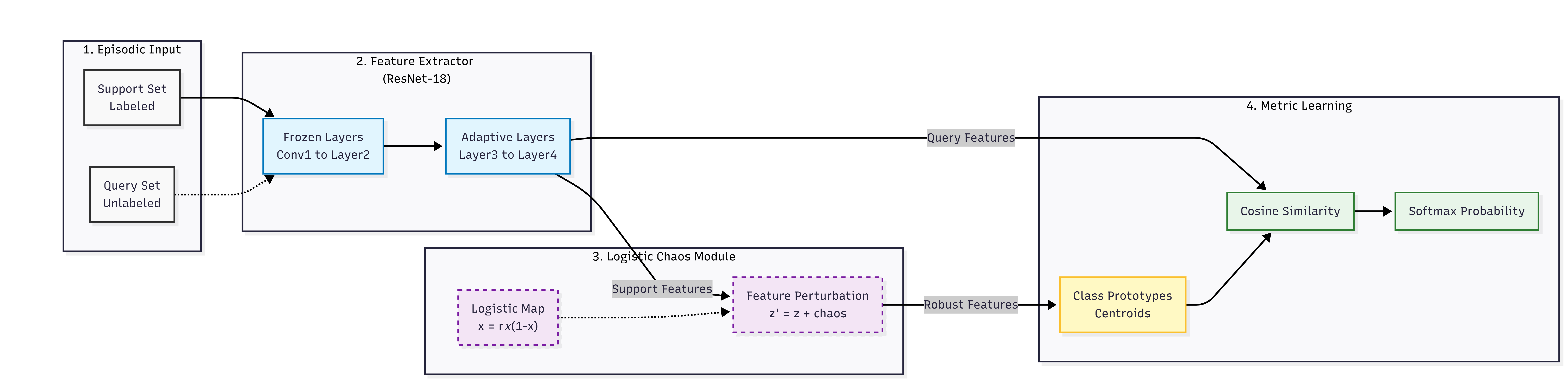}
    \caption{Architectural overview of the Chaos-Enhanced Prototypical Network (CE-ProtoNet). The framework Uses a pretrained ResNet-18 backbone with a \textbf{Selective Fine-Tuning} strategy (frozen early layers, trainable deep layers). The \textbf{Logistic Chaos Module (LCM)} introduces deterministic noise into the Support embeddings to create robust prototypes, while Query images bypass the chaos injection to ensuring deterministic inference.}
    \label{fig:architecture}
\end{figure}

\subsubsection{Hierarchical Feature Abstraction}
Medical imagery heavily relies on a hierarchy of visual information - where edges are domain-agnostic but lesion textures are highly domain-specific - we implement a \textbf{Selective Fine-Tuning Strategy}:
\begin{enumerate}
    \item \textbf{Frozen Layers (Conv1, Layer1, Layer2):} The initial blocks remain locked and they function as generic texture extractors, capturing the fundamental visual primitives that transfer cleanly from natural to medical images.
    \item \textbf{Adaptive Layers (Layer3, Layer4):} The final residual blocks stays fully unfrozen and makes the network to map complex semantic representations which are specific to neuro-oncology, such as the subtle boundary differences between different tumor types.
\end{enumerate}

\subsubsection{Embedding Projection}
A flattened vector $h \in \mathbb{R}^{512}$ is created from the output of the final convolutional layers which is processed through \textbf{Global Average Pooling (GAP)}. Recent studies \cite{11} have shown that the standard cosine distance frequently ignores the feature magnitude, often leading to create classification errors. To solve this we conducted strict \textbf{L2-Normalization}, projecting all embeddings onto a unit hypersphere:
\begin{equation}
z = \frac{f_\theta(x)}{||f_\theta(x)||_2 + \epsilon}
\end{equation}
 $\epsilon$ provides a numerical stability. This step helps to neutralize magnitude bias, making the subsequent metric depend on angular separation entirely \cite{11}.

\subsection{The Logistic Chaos Module (LCM)}
Techniques like Adversarial training or diffusion models can improve prototype fidelity, but they struggle with instability of batch-size \cite{17} and excessive computational complexity \cite{19} in few-shot environments. Our work introduces the \textbf{Logistic Chaos Module (LCM)} as a deterministic, computationally lightweight mechanism for injecting non-linear perturbations directly into the support embeddings on which we train to stabilize the embedding space.

\subsubsection{Dynamical System Definition}
The \textbf{Logistic Map} module is a discrete-time dynamical system governed by the recursive equation:
\begin{equation}
x_{n+1} = r \cdot x_n (1 - x_n)
\end{equation}
In the equation $x_n \in (0, 1)$ is the state variable. The bifurcation parameter is set to  $r = 3.99$. This setup guarantees a state of deep topological chaos. Earlier research have shown that logistic chaotic initialization maps local feature manifolds effectively than random stochastic noise, preventing convergence on localized optima \cite{18}.

\subsubsection{Chaos Injection Algorithm}
In episodic training process, the module applies the following operations to every support embedding $z_s$:
\begin{enumerate}
    \item \textbf{Initialization:} A random tensor is seeded and iterated through the logistic map for 8 ``warm-up'' cycles to eliminate transient correlations.
    \item \textbf{Centering and Scaling:} The resulting chaotic sequence $\delta$ is zero-centered and scaled by an intensity hyperparameter $\lambda$:
    \begin{equation}
    \text{Noise} = \lambda \cdot (2\delta - 1)
    \end{equation}
    \item \textbf{Perturbation:} The support vector is modified via addition:
    \begin{equation}
    \hat{z}_s = z_s + \text{Noise}
    \end{equation}
\end{enumerate}

To determine the appropriate perturbation intensity required we created an empirical sweep across multiple chaos intensities$\lambda \in \{0.0, 0.05, 0.10, 0.12, 0.15, 0.18, 0.20, 0.30, 0.40\}$. A clear trade-off between feature exploration and semantic destruction by the system. At low intensities ($\lambda \le 0.10$), the regularization was too weak, to prevent the network from memorizing static MRI artifacts, this caused overfitting. Similarly, when values exceeding $0.20$ distorted the embeddings, severely degrading prototype formulation and caused underfitting. 

The optimal threshold point was emerged at chaos intensity of $\lambda = 0.15$ ($15\%$). This intensity forced the network to explore the local feature manifold and discourages reliance on clinically irrelevant ``shortcut'' features \cite{13}, without removing the core disease profile. The LCM is kept disabled during validation and testing processes to ensure model benefits from chaotic regularization during training, clinical inference remains entirely deterministic and reproducible.

\subsection{Prototype Rectification and Metric Scaling}
The perturbed support set $\hat{S} = \{(\hat{z}_i, y_i)\}_{i=1}^{N \times K}$ is obtained once the deterministic noise is applied for the system. For a given class $c$, the \textbf{Rectified Class Prototype} $p_c$ is calculated from the perturbed embeddings centroid:
\begin{equation}
p_c = \frac{1}{K} \sum_{(x_i, y_i) \in S_c} \hat{z}_i
\end{equation}

This formula calculates the averages of dynamically ``jittered'' coordinates, resulting a smoothed centroid against the localized space irregularities. This helps the prototype from anchoring to the idiosyncratic noise of the $K$-shot samples, generating an invariant semantic core. These stabilized centroids then act as anchors for classifying unperturbed query images.

\subsubsection{Scaled Cosine Similarity}
Early stages of ProtoNets used a squared Euclidean distance, but instead using a scaled cosine similarity provides superior feature separability and reduces domain shift in high-dimensional medical tasks \cite{21}. To classify a query image $x_q$ with embedding $z_q$, we measure the scaled similarity against all prototypes using a learnable temperature parameter $\tau$ (initialized to $20.0$):
\begin{equation}
sim(z_q, p_c) = \tau \cdot \frac{z_q \cdot p_c}{||z_q||_2 \cdot ||p_c||_2}
\end{equation}

As the $z_q$ and $p_c$ are already $L_2$-normalized, this reduces to a simple dot product scaled by $\tau$. Unscaled cosine similarities are restricted within a range of $[-1, 1]$. Feeding this narrow range directly into a softmax function provides a low-confidence probability distribution \cite{22}. The parameter $\tau$ artificially expands this range. The sharper the softmax distributions are obtained using higher the values of $\tau$ . This actively penalizes the model for uncertain predictions and creating a tight intra-class clustering.

\subsubsection{Loss Function}
Optimization is performed for minimizing the negative log-probability of true class assignments for the query set. We use a cross-entropy loss over the scaled similarities:
\begin{equation}
\mathcal{L} = - \sum_{(x_q, y_q) \in Q} \log \left( \frac{\exp(sim(z_q, p_{y_q}))}{\sum_{k=1}^{N} \exp(sim(z_q, p_k))} \right)
\end{equation}

This orders a geometric layout of the latent space. While, the numerator maximizes metric alignment between a query $z_q$ and its ground-truth prototype $p_{y_q}$, the denominator forces the network to minimize similarity with the remaining incorrect prototypes. 

We obtain chaotic perturbed functions from the target prototype $p_k$, this loss function requires a very flexible mapping from the backbone encoder. This forces the network to compress the simulated tumor variations into dense local clusters while maximizing inter-class margins. This dual optimization reduces ambiguity between visually similar pathologies, and establishes the strict decision limits necessary for accurate clinical diagnosis.

\section{Experimental Results}
\label{sec:results}

To check the capability of the Chaos-Enhanced Prototypical Network (CE-ProtoNet) in data scarce regimes, we evaluated it on a public brain tumor MRI dataset. Our experiments were primarily focused on addressing three important questions that became central during our model development: (1) Can deterministic chaotic noise injection really provide better generalization than the standard ProtoNet? (2) What is the best mathematically optimal chaotic intensity ( $\lambda$ ) that encourages feature exploration without distorting the essential tumor profile? And (3), from a practical point of view, how will this regularized model handle visually challenging borderline cases? 

All the test's were conducted on  a 4-way 5-shot episodic few-shot protocol. The idea was to simulate a severely restricted clinical environmental settings, forcing the network to separate four distinct tumor classes using only five annotated reference scans per each class. For the feature extractor, we used a fine-tuned ResNet-18. Altough it is standard practice to utilize deeper architectures, but our preliminary results have shown that deploying models like ResNet-50 in such extreme low-data regimes caused immediate and severe overfitting \cite{10}. To isolate exactly how the Logistic Chaos Module (LCM) influenced learning, we conducted the network's performance across multiple injection intensities: $\lambda \in \{0.00, 0.05, 0.10, 0.12, 0.15, 0.18, 0.20, 0.30, 0.40\}$.

\subsection{Impact of Chaos Intensity (Ablation Study)}

\begin{figure}[h!]
\centering
\includegraphics[width=0.8\linewidth]{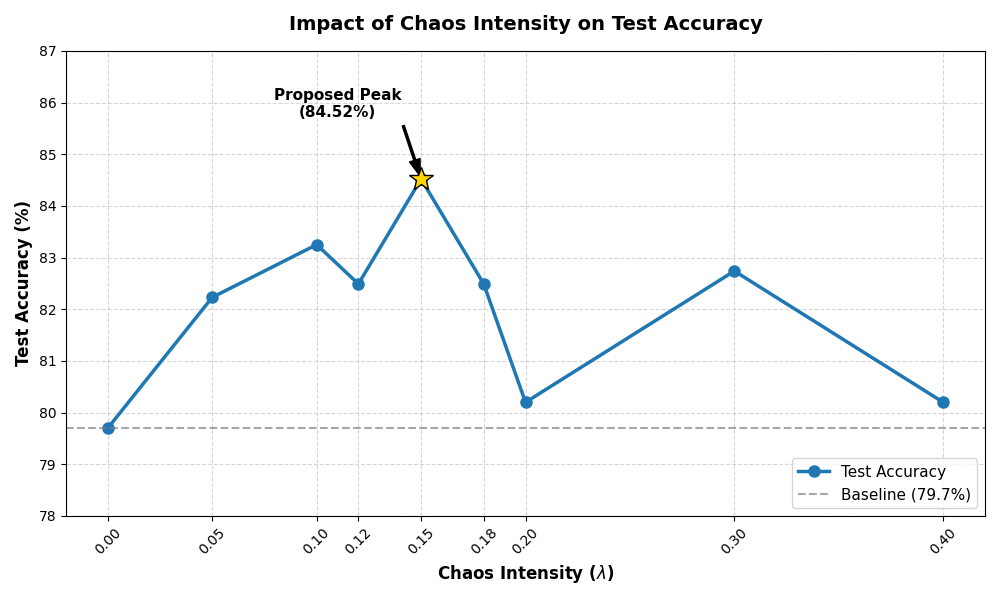}
\caption{Performance trajectory of test accuracy across an various chaos intensities.}
\label{fig:trend_graph}
\end{figure}

Experimentation was started by setting our base values ($\lambda=0.00$), and the standard prototype network struggled with the high intra-class variance settling at a test accuracy of 79.70\%. As injecting deterministic noise into the support set during training started, the resulting performance trajectory- detailed in Table \ref{tab:ablation} and Figure \ref{fig:trend_graph} - changed in quite obvious ways.
\begin{table}
\centering
\begin{small} % Slightly smaller text to ensure it fits the column width
\begin{tabular}{lcccc}
\toprule
\textbf{Chaos Intensity ($\lambda$)} & \textbf{Accuracy (\%)} & \textbf{Macro Precision} & \textbf{Macro Recall} & \textbf{Macro F1-Score} \\
\midrule
ProtoNet (Baseline) & 79.70 & 0.8683 & 0.7947 & 0.7665 \\
$\lambda = 0.05$ & 82.23 & 0.8745 & 0.8145 & 0.8058 \\
$\lambda = 0.10$ & 83.25 & 0.8847 & 0.8271 & 0.8190 \\
$\lambda = 0.12$ & 82.49 & 0.8805 & 0.8196 & 0.8088 \\
\textbf{$\lambda = 0.15$ (Proposed)} & \textbf{84.52} & \textbf{0.8915} & \textbf{0.8387} & \textbf{0.8354} \\
$\lambda = 0.18$ & 82.49 & 0.8819 & 0.8249 & 0.8047 \\
$\lambda = 0.20$ & 80.20 & 0.8608 & 0.7951 & 0.7823 \\
$\lambda = 0.30$ & 82.74 & 0.8693 & 0.8222 & 0.8164 \\
$\lambda = 0.40$ & 80.20 & 0.8681 & 0.7942 & 0.7834 \\
\bottomrule
\end{tabular}
\end{small}
\caption{Comparison of classification performance across varying Chaos Intensities ($\lambda$).}
\label{tab:ablation}
\end{table}

Introducing a mild chaos of ($\lambda=0.05$) created a sharp +2.53\% jump in test accuracy results. This observation has shown us that even low-level ergodic perturbation is enough to remove the model from sharp local minima, push it to generalize past the static visual noise of the limited $K$-shot samples.

Our experimental results have shown that this regularization effect hits a clear sweet spot at an intensity of $\lambda=0.15$. At this intensity, the model achieved a peak test accuracy of 84.52\%  and a macro F1 score of 0.8354, showing a net improvement of +4.82\% over standard ProtoNet. The macro recall has also peaked here (0.8387) showing that, this specific noise threshold successfully captures intra-class variance without forcing features into pathological overlap.

But pushing the perturbation beyond this optimal point has caused the feature space begin to structurally deform. At $\lambda \ge 0.18$ we started to see a decrease in the results, accuracy started falling to 80.20\% at both 0.20 and 0.40 chaos intensities. An unusual upturn was observed at $\lambda=0.30$ (82.74\%) which we suspect as the byproduct of stochastic resonance  which has temporarily induced the embedding in a favorable geometry, but the broader trend was undeniably positive. Excessive noise overwrites the intrinsic biological identifiers of tumor sections, destroying the network's ability to generate specific prototypes.

\subsection{Error \& Class-Wise Analysis}

\begin{figure}[h!]
\centering
\begin{subfigure}{0.48\textwidth}
\centering
\includegraphics[width=\textwidth]{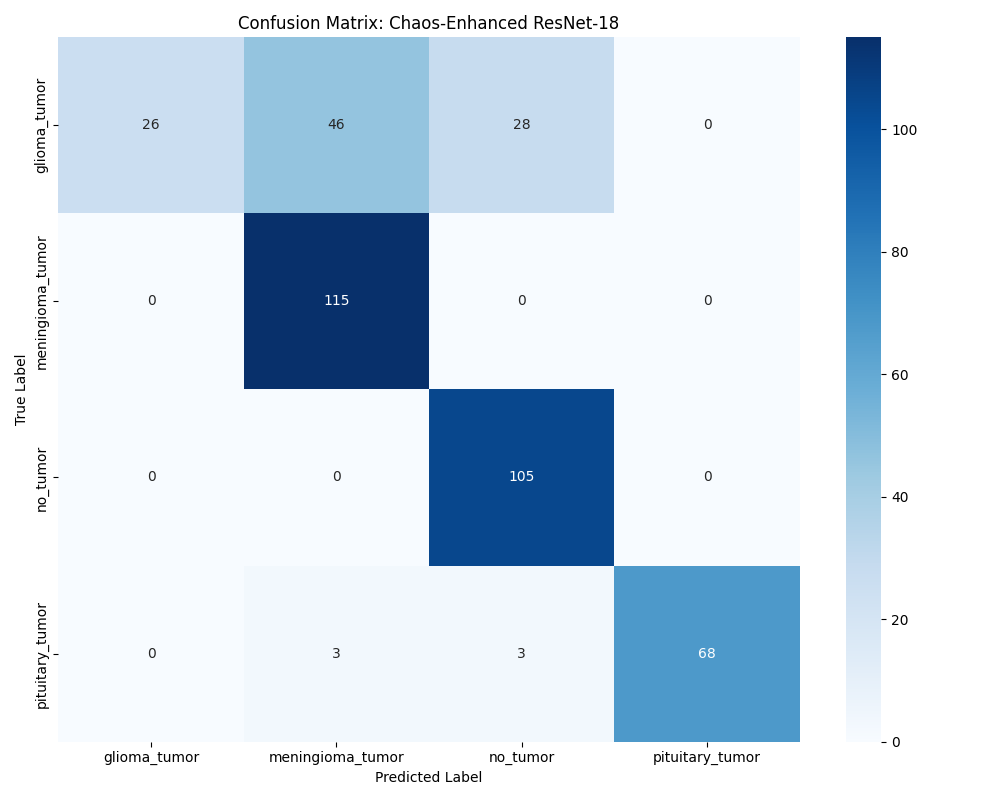}
\caption{ProtoNet ($\lambda = 0.00$)}
\label{fig:cm_baseline}
\end{subfigure}
\hfill
\begin{subfigure}{0.48\textwidth}
\centering
\includegraphics[width=\textwidth]{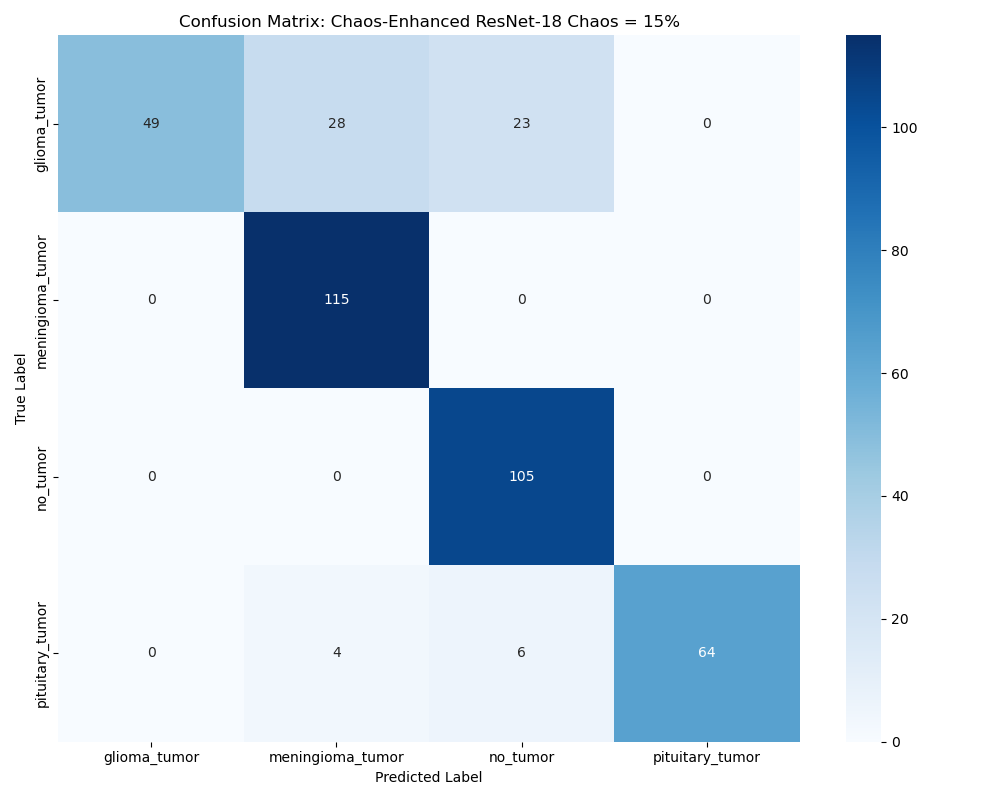}
\caption{Proposed ($\lambda = 0.15$)}
\label{fig:cm_proposed}
\end{subfigure}
\caption{Comparative confusion matrices demonstrating the regularizing effect of the Logistic Chaos Module.}
\label{fig:cm_comparison}
\end{figure}

To understand the behavior of this framework, we broke down the diagnostic performance by class. We compared its confusion matrices of ProtoNet against the optimal $\lambda=0.15$ point (Fig. \ref{fig:cm_comparison}), and analyzed the detailed metrics in Table \ref{tab:classwise}.

\begin{table}[h!]
\centering
\begin{tabular}{lcccc}
\hline
\textbf{Class Label} & \textbf{Precision} & \textbf{Recall} & \textbf{F1-Score} & \textbf{Support Samples} \\
\hline
Glioma Tumor & 1.0000 & 0.4900 & 0.6577 & 100 \\
Meningioma Tumor & 0.7823 & 1.0000 & 0.8779 & 115 \\
No Tumor & 0.7836 & 1.0000 & 0.8787 & 105 \\
Pituitary Tumor & 1.0000 & 0.8649 & 0.9275 & 74 \\
\hline
\end{tabular}
\caption{Detailed class-wise performance metrics for the optimal CE-ProtoNet configuration ($\lambda=0.15$).}
\label{tab:classwise}
\end{table}

The key-point from this breakdown was the recovery in Glioma recall. Gliomas are heavily difficult to classify due to their characteristics like diffuse margins and extreme morphological heterogeneity. In our ProtoNet tests, the static prototypes for Gliomas essentially collapsed, heavily overlapping with adjacent classes predominantly Meningiomas. This resulted in a lower recall of approximately 0.26. 

Implementing the 15\% chaotic perturbation, we have simulated a natural biological variance during training. This forced the network to learn boundary-invariant representations, nearly doubling the Glioma recall to 0.4900. While this remains the lowest recall metric across the dataset when comparted to other classes, but the intervention clearly halted the complete marginalization of these diffuse features. 

The model got a precision of 1.0000 for gliomas. This results shows that CE-ProtoNet learns to generate a highly conservative, tightly bound prototype for this specific class. This limitation directly affects the characteristics of meningioma tumor. Similar to the glioma class, the meningioma also achieved a perfect 1.0000 recall, but a relatively lower 0.7823 precision. As the glioma threshold is so tight, the model captures all true meningiomas, but ends up absorbing highly scattered, ambiguous glioma samples including false negatives as false positives. This trade-off fully shows the inherent visual ambiguity between these two classes of MRI modalities that lack contrast enhancement. Chaos successfully helped in separating feature clusters, but residual overlap mainly favors mean prediction.

Importantly, we also observed excellent semantic consistency across the other classes. Adding randomness can accidentally ruin the most important signals that help a system make accurate distinctions. However, the CE-ProtoNet efficiently handled morphologically distinct classes exceptionally well. The pituitary tumor class achieved an F1 score of 0.9275, supported by perfect precision (1.0000) and high recall (0.8649). Because these tumors occur in a highly localized, specific anatomical region, chaotic injection has effectively tightened the cluster without spreading to adjacent functional sites. The No Tumor class similarly retained an ideal 1.0000 recall, giving us assurance that the empirically derived $\lambda=0.15$ threshold works as a non-destructive regularizer solving high-variance edge cases without sacrificing high-frequency identifiers of specific, healthy tissue.

\subsection{Geometric and Qualitative Validation}

\begin{figure}[h!]
\centering
\includegraphics[width=1.0\linewidth]{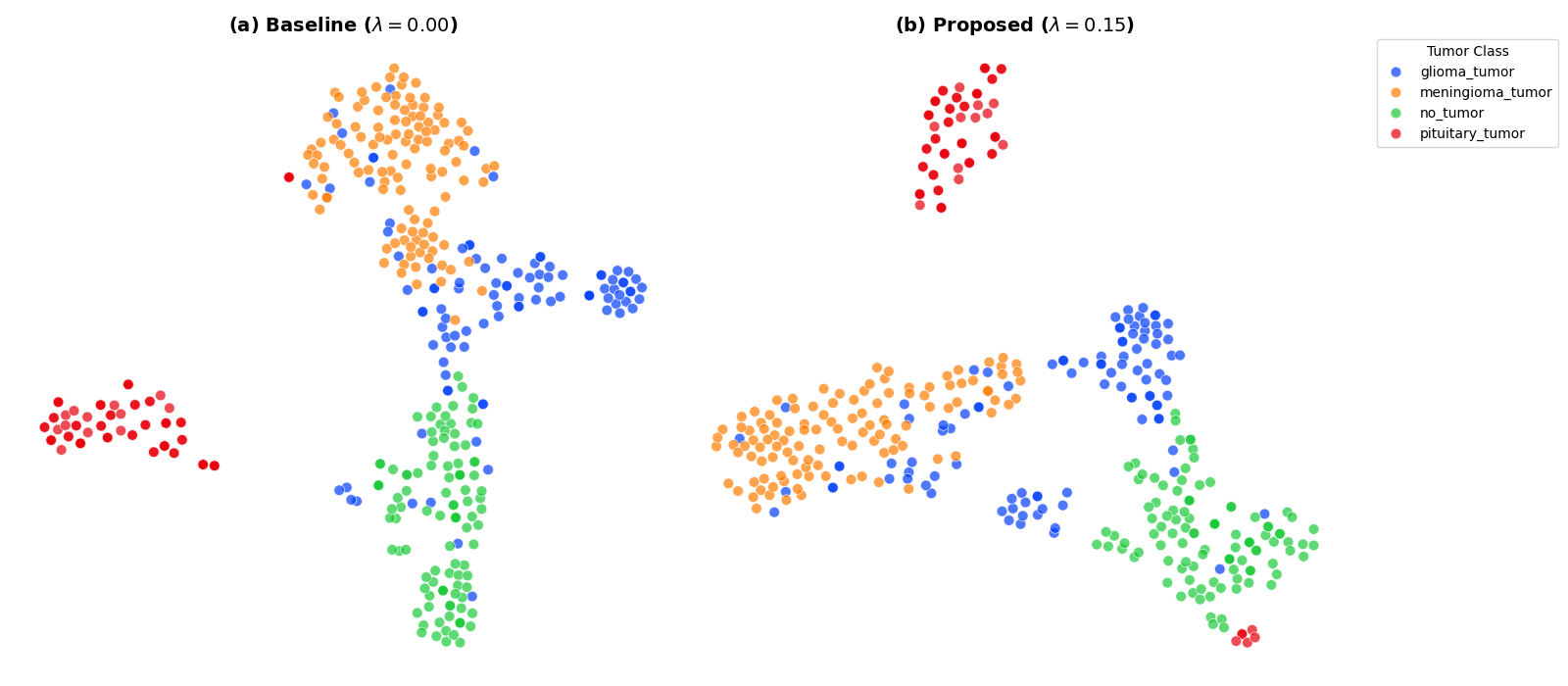}
\caption{t-SNE visualization of feature embeddings on the Test Set. \textbf{(a) ProtoNet ($\lambda=0.00$):} The projection reveals severe entanglement between Glioma (Blue) and Meningioma (Orange). \textbf{(b) Proposed ($\lambda=0.15$):} The injection of deterministic chaotic noise geometrically reorganizes the manifold, forcing cluster centroids apart and tightening intra-class variance.}
\label{fig:tsne_comparison}
\end{figure}

To verify what we were seeing in the classification metrics in a geometrical prospect, we projected the test set embeddings into a 2D plane using t-SNE (Fig. \ref{fig:tsne_comparison}). The two models look quiet different from each other.

In the ProtoNet projection (Fig. \ref{fig:tsne_comparison}a), the high-dimensional representations of Glioma (blue) and Meningioma (orange) samples are severely entangled. This manifold visualizes why we saw near-total feature collapse in the ProtoNet's metrics; the prototype was anchored in a region saturated with overlapping class noise. Likewise, the CE-ProtoNet projection (Fig. \ref{fig:tsne_comparison}b) shows a reorganized and stabilized space. Injecting a continuous 15\% chaotic noise appears to act as an active spatial repellent. While some semantic proximity naturally arise between Gliomas and Meningioma classes, their respective cluster centroids are moved further apart, carving out a distinct inter-class margin. Furthermore, the clusters for the pituitary and no tumor classes are closely packed around their prototypes, confirming the drastic reduction in intra-class variance. This geometric evolution provides strong evidence that ergodic perturbations prevent the backbone encoder from settling into overfitted subspaces.

\begin{figure}[h!]
\centering
\includegraphics[width=1.0\linewidth]{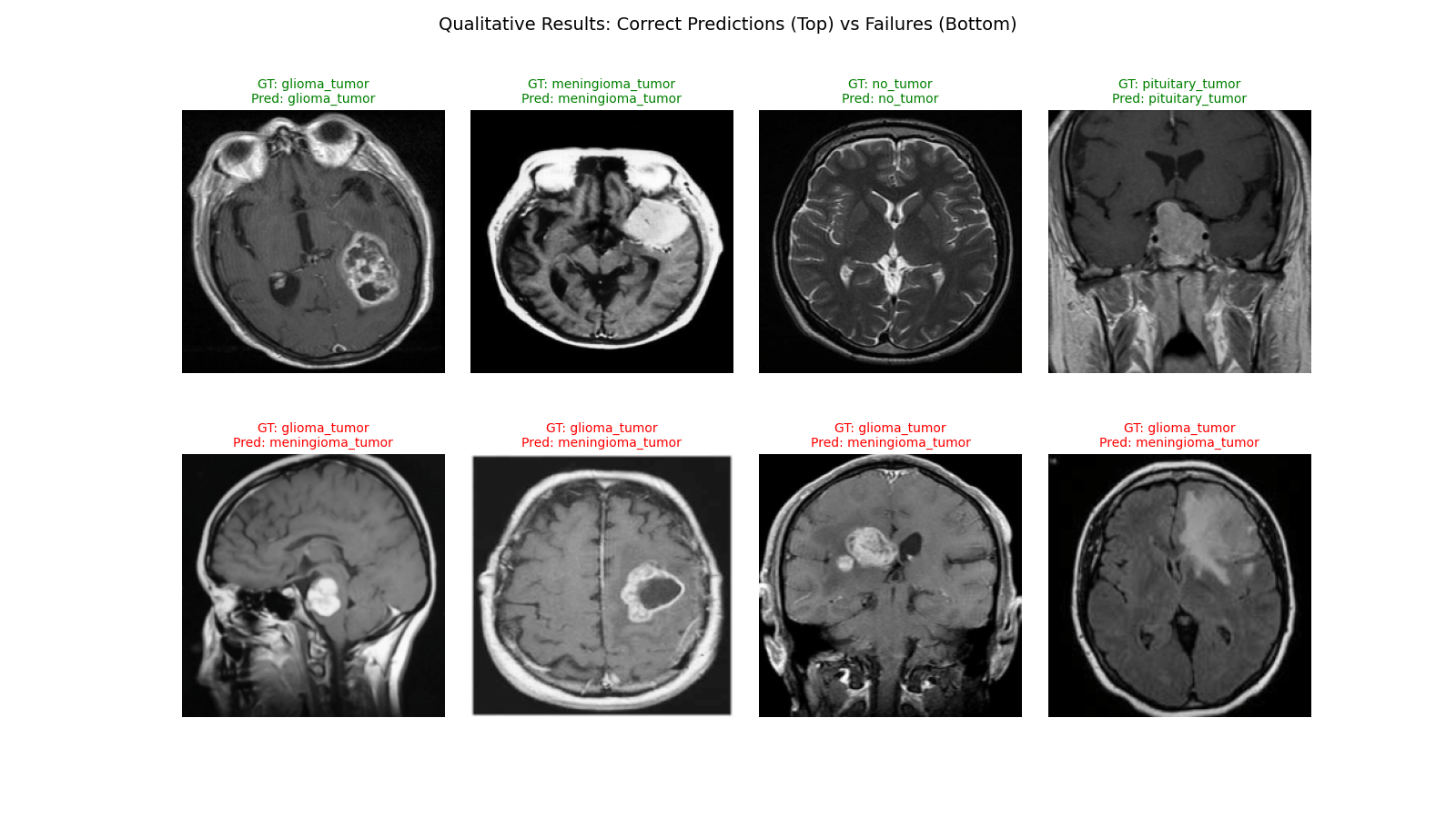}
\caption{Qualitative assessment of model predictions. \textbf{Top Row (Successes):} Correct classifications. \textbf{Bottom Row (Failures):} The model struggles with diffuse Gliomas, frequently misclassifying them as Meningiomas.}
\label{fig:qualitative}
\end{figure}

Finally, we performed a visual review of random test samples to contextualize these findings clinically (Fig. \ref{fig:qualitative}). 

When the model is successful (Fig. \ref{fig:qualitative}, top row), it shows an excellent ability to anchor true structural anomalies rather than being dominated by background noise, which was particularly evident in the pituitary class case. However, the failure cases (Fig. \ref{fig:qualitative}, bottom row) consistently are from glioma-meningioma overlap. When encountering diffuse gliomas that lack a clear defining feature for example, a prominent necrotic core—the model often mislabels them as meningiomas because their pixel intensity and visual texture are similar. 

In the absence of explicit internal markers, the network is forced to rely on peripheral boundary features, which naturally overlap between the two classes. While chaotic regularization successfully separated most classes and improved the overall clinical scenario, fully resolving these highly ambiguous edge cases seems beyond the reach of single-modality MRI. Based on these observations, it is clear that future work will need to integrate multimodal MR data, for example by combining T1-weighted contrast and T2-FLAIR sequences to provide the tissue contrast necessary to eliminate these specific misclassifications.
   
\section{Discussion}
\label{sec:discussion}

When we first started evaluating CE-ProtoNet, our primary question was whether deterministic chaotic dynamics could really regularize a network in medical imaging scenarios with very little data. We found that the architecture increased our peak diagnostic accuracy to $84.52\%$ , which is a remarkable improvement compared to the ProtoNet ($79.70\%$ ). In this section, we want to explain why we believe this geometric stabilization occurs, how it succeeds in keeping the computational overhead low compared to current standard practice, and how single-sequence MRI fundamentally limits the classifier.

\subsection{Mechanism of Improvement: Rectifying Prototype Bias}
One of the most critical issues we noticed during early ProtoNet testing was ``prototype bias'' \cite{12}. With so few support samples per episode, the empirical centroid of a class often shifts significantly from the true distribution of the class. The ProtoNet accuracy of $79.70\%$ (Table \ref{tab:ablation}) essentially reflects the severity of this geometric drift. 

By introducing the Logistic Chaos module with a fixed intensity ($\lambda=0.15$) appears to force a spatial enhancement. Initially, we tried to regularize the latent space with standard Gaussian noise, but its unbounded nature was often too destructive and collapsed class clusters altogether. However, the logistic map produces a trajectory which is deterministic, tightly bounded, and ergodic.This mathematical distinction proved critical in our experiments. Instead of computing a class prototype from a few isolated coordinate points, the network tries to find the stable center of a comprehensively perturbed volume. This aligns with theoretical work suggesting that chaotic sequences actively keep clustering algorithms out of sharp local minima \cite{18}. In our setting, this ergodic expansion seemed to help the network capture the severe intra-class variance we saw in diffuse Gliomas—variance that our static ProtoNets systematically missed.

\subsection{Architectural Efficiency vs. State-of-the-Art}
Recent studies, such as \textit{ProtoMed} \cite{10}, depends heavily on auxiliary contrastive losses to enforce tight cluster margins. We experimented with similar hard-negative mining techniques in early stages, but we found them computationally complex and heavily unstable across the small batch sizes which are required for episodic training. CE-ProtoNet changes everything by driving the spatial regularization strictly through forward-pass chaotic perturbation, we have achieved a similar tightening of the clusters across Pituitary and No Tumor classes, as seen in Fig. \ref{fig:tsne_comparison} without the fragility of balancing complex loss terms.

We specifically designed this module such that it avoids the overhead of generative augmentation. Techniques like Diffusion Models (DDPMs) and GANs are incredibly popular right now for synthesizing medical data \cite{17, 19}, but they require heavy computation to generate. More importantly, they introduce a real clinical risk of hallucinating and creating anatomically impossible features. Because our Logistic Chaos Module operates with strict $\mathcal{O}(1)$ algorithmic complexity, it simulates the regularizing effect of heavy data augmentation directly in the latent space, without the prohibitive costs or risks of pixel-level generation. This keeps the architecture effective for parameter-efficient learning \cite{20}.

\subsection{The Glioma-Meningioma Modality Trap}
The overall accuracy metrics show some persistent class-specific struggles. Looking at the granular data in Table \ref{tab:classwise}, Gliomas and Meningiomas remain a major bottleneck. The injection of chaotic perturbation successfully pushed Glioma precision to $1.0000$ (effectively eliminating false positives in our test sets), and recall improved from $0.26$ to $0.49$. Still, a recall of $0.49$ is obviously constrained. 

On reviewing the qualitative assessments (Fig. \ref{fig:qualitative}), it has shown us that this isn't necessarily a failure of the algorithm, but an inherent limitation of the imaging modality itself. In non-contrast T1-weighted MRI sequences, diffuse Gliomas and Meningiomas frequently share nearly identical textural profiles which raises the ambiguity. Without multiparametric input, such as T2-FLAIR or post-contrast sequences where the anatomical boundaries separating these tissues lack the necessary gradient contrast. As basic metric learning principles dictate, a distance-based classification cannot resolve boundary ambiguity if the source embeddings are structurally synonymous. We can mathematically filter out spurious noise, but algorithmic regularization cannot spontaneously generate physiological textures that are physically missing from the scan.

\subsection{Limitations and Future Trajectories}
There are also few clear limitations on how we apply the chaotic noise. The current iteration of CE-ProtoNet depends on a static value, which is globally applied as chaos intensity (Example: $\lambda=0.15$). While our ablation study shows that this is the best compromise across the whole dataset, we did observe a slight drop in Meningioma precision compared to runs with lower injection intensities. This shows that applying a uniform perturbation is sub-optimal; highly heterogeneous pathologies clearly have different mathematical tolerances for feature jitter.

Consequently, our future directive is to move away from a static parameter and develop an Adaptive Chaos Module. We hypothesize that if we make $\lambda$ a dynamically learnable coefficient-perhaps conditioned on the spatial entropy or predictive uncertainty of the specific input embedding \cite{8}-the network could selectively hit highly ambiguous samples with aggressive ergodic exploration while leaving distinct, high-confidence morphologies alone. Combining this adaptive intensity with local attention mechanisms seems like a very practical path to pursue parameter-efficient few-shot ontologies, and it forms the basis of our ongoing work.

\section{Conclusion}
\label{sec:conclusion}

In this work, we investigated how to solve the severe overfitting in which we routinely observed when training Prototypical Networks on sparse medical imaging data. Initially, our attempts to regularize the network with normal dropout or gaussian noise did not lead to much improvement in the stabilization of the latent space geometry. This resulted in the design of the Chaos-Enhanced Prototypical Network (CE-ProtoNet). It has been observed that by inserting a Logistic Chaos Module into a fine-tuned ResNet-18, we actively perturbed the embeddings during the episodic training. We have chosen this deterministic way in order to encourage the network to explore the feature space more than random noise without increasing the total number of parameters in the model.

Evaluating this set-up with a brain tumor MRI dataset, we found that it is sensitive to the intensity of the chaotic perturbations. By determining the injection parameter $\lambda$ to about 0.15, we found that it enabled the best empirical balance, and last but not least, our test accuracy was increased from 79.70\% to 84.52\%. In our setting, limiting the perturbation through the logistic map stabilized the functions better than in our previous experiments done with unstructured random noise several times better to the point that completely perturbed class clusters in these data poor regimes were quite common.

However, not all tumor types performed well with this type of therapy. The model worked well for anatomically distinct classes - we found an F1-score of 0.9275 in pituitary tumors - however, the drawback of single modality non-contrast T1 sequences was evident when parsing highly diffuse tissues. A look at, for instance, the chaotic regularization showed that, although the chaotic regularization succeeds in bringing gliomas closer to 1.0000 (thus effectively eliminating false positives for our test splits), this is doing so at a serious cost to recall (which then falls to 0.4900). The latent embeddings for gliomas and meningiomas were entangled nonetheless as visualized. We suspect this trade-off reveals a hard limit on what can be separated without contrast enhancement or multi-modal inputs, as opposed to an issue with the metric learning approach itself. 

From a practical point of view, we have created this framework so as not to cause the computational overhead of creating synthetic medical data using GANs or diffusion models. As the Logistic Chaos module runs at a $\mathcal{O}(1)$ complexity, the training costs is almost the same as used in the standard ProtoNet. This lightweight profile facilitates introduction to clinical diagnostics setting which are often limited by legacy hardware.

Given residual entanglement between more dispersed tumor sections, depending on the intensity of static perturbations in all of the episodes is likely to be suboptimal. An obvious next step to this research is to make chaotic intensity ($\lambda$) as an adaptive Feature. As mentioned above, if we can control the intensity of chaos is on entropy of current batch, the network can theoretically only use the aggressive regularization, in case of highly ambiguous in front support samples, except high confidence embeddings. Overall, the combination of discrete dynamical systems and deep metric learning seems to be a practical yet parameter-efficient approach to address the data scarcity problem in neuro-oncology, and our tests with this adaptive formulation are planned in our ongoing work.

\section*{Declarations}

\textbf{Funding} \\
The authors declare that no funds, grants, or other support were received during the preparation of this manuscript.

\textbf{Competing Interests} \\
The authors have no relevant financial or non-financial interests to disclose.

\textbf{Data Availability} \\
The Brain Tumor MRI dataset analyzed during the current study is available in Kaggle, accessible via: \url{https://www.kaggle.com/datasets/sartajbhuvaji/brain-tumor-classification-mri/data}.

\textbf{Code Availability} \\
The source code for the Chaos-Enhanced Prototypical Network, including the implementation of the Logistic Chaos Module and the pre-trained weights, is available at: \url{https://github.com/meghan-reddy6/xxxxxxxxxx}.

\textbf{Ethical Approval} \\
This article does not contain any studies with human participants or animals performed by any of the authors. The dataset used is publicly available and anonymized.

\end{document}